\documentclass[showpacs,twocolumn,aps,floatfix,superscriptaddress]{revtex4}
\usepackage{amsmath,amssymb,eucal,graphicx,subfigure}
\begin{document}

\def\a{\alpha}
\def\b{\beta}
\def\P{\hat P}
\def\F{\hat F}
\def\av#1{\langle#1\rangle}

\title{Asymptotic behavior of the Kleinberg model}

\author{Shai Carmi}
\affiliation{Minerva Center \& Department of Physics,
Bar-Ilan University, Ramat Gan 52900, Israel}
\affiliation{Center for Polymer Studies, Boston University, Boston,
MA 02215, USA}
\author{Stephen Carter}
\affiliation{Department of Mathematics \& Computer Science, Clarkson University, Potsdam, NY 13699-5815}
\author{Jie Sun}
\affiliation{Department of Mathematics \& Computer Science, Clarkson University, Potsdam, NY 13699-5815}
\author{Daniel ben-Avraham}
\email{benavraham@clarkson.edu}
\affiliation{Physics Department, Clarkson University, Potsdam, New York
13699-5820, USA} 
\affiliation{Department of Mathematics \& Computer Science, Clarkson University, Potsdam, NY 13699-5815}

\begin{abstract}
We study Kleinberg navigation  (the search of a target in a $d$-dimensional lattice, where each site is connected to  one other random site at distance $r$, with probability $\sim r^{-\a}$) by means of an exact master equation for the process.   We show that the asymptotic scaling behavior for the
delivery time $T$ to a target at distance $L$ scales as $T\sim\ln^2L$ when $\a=d$, and otherwise as $T\sim L^x$, with $x=(d-\a)/(d+1-\a)$ for
$\a<d$, $x=\a-d$ for $d<\a<d+1$, and $x=1$ for $\a>d+1$.  These values of $x$ exceed the rigorous lower-bounds established by Kleinberg.  We also address the situation where there is a finite probability for the message to get lost along its way and find short delivery times (conditioned upon arrival) for a wide range of $\a$'s.
\end{abstract}

\maketitle

In a  now famous study~\cite{milgram} the social psychologist Stanley Milgram asked randomly chosen people
to send a postcard to a disclosed target in the USA.  The participants were to send the card to a person they knew on a first-name basis, who will then send it on to another acquaintance, and so on, until it reached the  desired destination.  20\% of the cards successfully reached the target using, on average, chains of $6.5$ acquaintances, confirming the notion that the network of social contacts has the {\it small-world\/} 
property~\cite{watts}: a very short path, typically logarithmic in the size of the system, connects between any two nodes.
How does the message find its way, let alone so efficiently?  Searching and navigation problems such as this~\cite{kleinberg1,kleinberg2,boguna,lomholt,oshanin,benichou,edwards,chen,santos,adamic,watts1}
are relevant to several disciplines, from sociology, to efficient algorithms in computer science, to the understanding of the foraging behavior of insects and animals.  

The problem has been elegantly addressed in the seminal work of Jon Kleinberg~\cite{kleinberg1,kleinberg2}.
Kleinberg considers an $(L\times L)$-square lattice,  where in addition to the links between nearest neighbors each node $i$ is connected to a random node $j$ with a probability  
$p_{ij}=r_{ij}^{-\alpha}/\sum_k r_{ik}^{-\alpha}$ ($r_{ij}=|{\bf r}_i-{\bf r}_j|$ is the Euclidean distance between nodes $i$ and $j$, and the sum in the denominator excludes  $k=i$).  Suppose that a message is to be passed from a ``source" node $s$ to a ``target" node $t$, along the links of the network, by a {\it decentralized\/}  algorithm --- an algorithm that relies solely on {\it local\/} information.  Kleinberg shows that when the exponent $\alpha$ equals $d$, the lattice dimensionality, an algorithm exists that requires { less} than $\ln^2 L$ steps to complete the task.  If $\alpha\neq d$, the delivery time, $T$, grows as $L^x$, with rigorous
lower bounds, $x\geq x_K$~\cite{kleinberg1,kleinberg2,roberson}; 
\begin{equation}
\label{x_K}
T\sim L^x\,,\qquad x\geq x_K=
\begin{cases}
\frac{d-\a}{d+1} & 0\leq\a<d\,,\\
\frac{\a-d}{\a-d+1} & \a>d\,.
\end{cases}
\end{equation}
Moreover, no local algorithm performs better, functionally, than the simple-minded {\it greedy\/} algorithm:
{\it Pass the message forward to the neighbor node that is closest to the target\/} (geographically).

In this letter we study the asymptotic long-time behavior of the Kleinberg search process.
We find the exact form of the exponent $x(\a)$, and we show that $T\sim\ln^2L$ is the actual scaling (not just a bound)
for the special case of $\a=d$.  Our approach is based on a master equation for the
full probability distribution for completing a search within a given time.  This formalism also enables one
to consider the possibility of the message getting lost along its way, and we discuss briefly
some surprising outcomes of that scenario.

Because the message gets closer to the target with each successive step, nodes are never revisited and one can
view each long-range link as being created only as the message arrives at its site of origin.  The message closes in on the target in a peculiar kind of directed L\'evy walk, consisting of a mix of ``short" (one lattice spacing) and ``long" (power-law distributed) steps.  In this search the dimensionality of the lattice enters into consideration mainly as a Jacobian in the various sums (or integrals) of the equations describing the process.  We therefore limit the following derivations to
one dimension and generalize the results for higher dimensions, having made the necessary adjustments.

For convenience, to render the $p_{ij}$ independent of $i$, we adopt periodic boundary conditions.
Specifically, consider a ring of length $4L$, with the source at $0$ and the target at $L$, and the range of the long-contact
links limited to $2L-1$~\cite{remark1}.  In that case,
\begin{equation}
\label{pijA}
p_{i,i+k}=A k^{-\a}\,;\qquad A=\Big(2\sum_{k=1}^{2L-1}k^{-\a}\Big)^{-1},
\end{equation}
where $A$ is a normalizing factor.
Let $P(n;l)$ denote the probability that a message, at distance $l$ from the target, takes $n$ additional steps
to reach the target.  Once the message is at the target it takes
no additional time to reach it, so $P(n;0)=\delta_{n,0}$.  Likewise, $P(0;l)=\delta_{0,l}$, since the only way to reach the target
in 0 steps is if the message is already there to begin with.

$P(n;l)$ satisfies the equation
\begin{equation}
\label{Pnl}
\begin{split}
P(n;l)=A&\sum_{k=1}^{2l-1}k^{-\a}P(n-1;|l-k|)\\
&+\Big(1-A\sum_{k=1}^{2l-1}k^{-\a}\Big)P(n-1;l-1)\,.
\end{split}
\end{equation}
The first term on the rhs represents the events that the first of the additional $n$ steps is a long step of length $k$,
in which case the message would come to within distance $|l-k|$ from the target.  The second term represents
a short step, that advances the message a single lattice spacing.  

\begin{figure}[]
\includegraphics[width=0.5\textwidth]{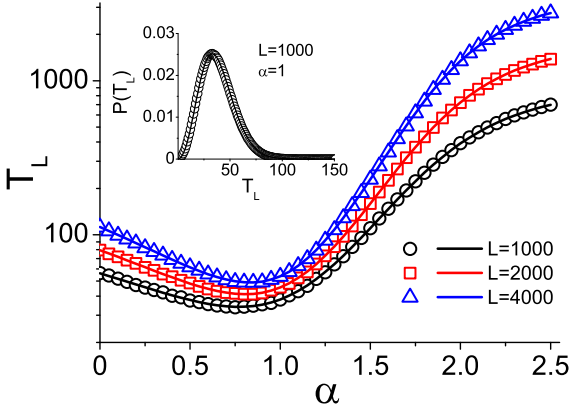}
\caption{(Color online) Mean delivery time, $T_L$, as a function of the long-contact exponent, $\alpha$.
Note the perfect agreement between simulations (solid line) and the results from Eq.~(\ref{Tl}) (symbols).
Shown are results for three values of $L$.  Inset: Distribution of the delivery time for the case of $\a=1$ and $L=1000$ as computed from~(\ref{Pnl}) (solid line) is compared to simulations (symbols). }
\label{T-alpha}
\end{figure}

For our main purpose here it is sufficient to consider just the first moment $\av{n}\equiv T_l$, that is, the mean delivery time from a site a distance $l$ away from the target. Multiplying Eq.~(\ref{Pnl}) by
$n$  and summing over $n$, we get
\begin{equation}
\label{Tl}
\begin{split}
T_l=A&\sum_{k=1}^{2l-1}k^{-\a}\left(1+T_{|l-k|}\right)\\
&+\Big(1-A\sum_{k=1}^{2l-1}k^{-\a}\Big)\left(1+T_{l-1}\right)\,,
\end{split}
\end{equation}
for $l=1,2,\dots,L$.  Numerical integration of Eqs.~(\ref{Pnl}) and (\ref{Tl}) yields perfect agreement
with the results from direct simulation of the Kleinberg navigation process on a ring (Fig.~\ref{T-alpha}).

Using the fact that $T_{-k}=T_k$, and defining $D_k=T_k-T_{k-1}$,
we obtain, after some rearranging,
\begin{equation}
\label{dD}
\begin{split}
D_{l+1}-D_l=A\Big\{&\sum_{k=1}^l\big[(l+1-k)^{-\a}\\
&-(l+k)^{-\a}\big]D_k-\sum_{k=1}^{2l-1}k^{-\a}D_l\Big\}\,,
\end{split}
\end{equation}
for $l=1,2,\dots,L-1$, while for $l=0$ we have $D_1=T_1-T_0=1$.

As a quick check, consider the limit of
$\a\to\infty$, when all the long-range contacts are restricted to length 1, and therefore one expects $T_l=l$.
Indeed, in this case all the $k^{-\a}$ terms in the equation tend to zero, unless $k=1$, and we get $D_{l+1}-D_l=0$, which along with $D_1=1$ yields $D_k=1$, and $T_l=\sum_{k=1}^lD_k=l$, just as expected.

Next, consider the opposite limit, of
 $\a=0$, where the distribution of long-rage contacts is homogeneous.  In this case $A=[2(2L-1)]^{-1}$ and we obtain from (\ref{dD}),
\[
D_{l+1}-D_l=-\frac{2l-1}{2(2L-1)}D_l\,.
\]
Although this equation can be solved exactly, a continuous approximation, 
\[
\frac{d}{dl}D(l)=-\frac{l}{2L}D(l)\,,
\]
assuming $L\geq l\gg1$, works just as well.
In view of the boundary condition $D(0)=1$, this has the solution $D(l)=\exp(-l^2/4L)$.  Then,
$T(L)=\int_0^LD(l)\,dl$. The upper integration limit may be safely replaced with $\infty$, due to the rapid decay of the gaussian, and a simple change of variables yields $T(L)\sim L^{1/2}$,
in perfect agreement with the Kleinberg bound for $\a=0$.

For larger values of $\a$ we are not as fortunate as to find a full analytic solution, but we can still obtain the asymptotic behavior.  For $0\leq\a<1$ we take a hint from the solution for $\a=0$ and
 make the {\it ansatz\/}  $D(l)=f(l^{\beta}/L)$, where $f(x)$ is  a smoothly decreasing function; $f(x)={\cal O}(1)$ for $x\lesssim1$, and decays very rapidly (e.g., exponentially) for $x\gtrsim1$.  Consistent with this behavior, the derivative at the crossover point $x_*=1$ is $f'(x_*)=-{\cal O}(1)$.  
This ansatz is nicely confirmed by numerical integration of Eq.~(\ref{dD}).

Apply now Eq.~(\ref{dD}) to the crossover length $l_*=L^{1/\beta}$.  The lhs is 
\[
D_{l_*+1}-D_{l_*}\approx\frac{d}{dl}D(l)|_{l=l_*}\sim -L^{-1/\beta}\,,
\]
while the sums on the rhs can be estimated by replacing $D_l$ with a constant for $l<l_*$, and zero for $l>l_*$, yielding $ -Al_*^{1-\a}$.  But $A\sim1/L^{1-\a}$, leading to $-1/\beta=(1-\a)/\beta-(1-\a)$, or
$\beta=(2-\a)/(1-\a)$.  Finally,
\[
\begin{split}
T(l)=\int_0^L&D(l)\,dl\approx\int_0^{\infty}D(l)\,dl\\
&=\frac{L^{1/\beta}}{\beta}\int_0^{\infty}f(x)x^{1/\beta-1}\,dx\sim L^{1/\beta}\,,
\end{split}
\]
so 
\begin{equation}
T(L)\sim L^{(1-\a)/(2-\a)}\,,\qquad 0\leq\a<1\,.
\end{equation}

For $\a>1$, 
we sum Eq.~(\ref{dD}) over $l$, taking into account that $D_1=1$, and rearrange:
\begin{equation}
\label{D1}
D_{l+1}-1=-A\sum_{k=1}^l\sum_{m=-k+2}^k(l+m)^{-\a}D_k\,.
\end{equation}
This can be obtained more directly also by rearranging~Eq~(\ref{Tl}). 
The inner sum over $m$ can be approximated 
by an integral, yielding
\begin{equation}
\label{D1c}
D_{l+1}-1\approx\frac{A}{\a-1}\sum_{k=1}^l\left[(l+k)^{1-\a}-(l-k+2)^{1-\a}\right]D_k\,.
\end{equation}
Since $\a>1$, $A$ converges, as $L\to\infty$, and we may
simply follow powers of $l$.
Assume first that $D_l\sim l^{-\beta}$.
The dominant $-1$ on the lhs of the equation must be balanced by the dominant term on the rhs, which scales as $-l^{2-\a-\beta}$,
so $\beta=2-\a$.  This leads to 
\begin{equation}
T\sim L^{\a-1}\,,\qquad1<\a<2\,.
\end{equation}  
The upper limit on $\alpha$ follows from the fact that $T_L$ cannot increase faster than linearly in $L$.

For the special case of $\alpha=1$, we have
\[
D_{l+1}-1\approx{A}\sum_{k=1}^l\ln\frac{l+k}{l-k+2}D_k\,,
\]
and $A\sim(\ln L)^{-1}$. To counter the $-1$ on the lhs one must then allow that $D_l\sim (\ln L)/l$, for large $l$,
which leads to 
\begin{equation}
T\sim\ln^2L\,,\qquad \a=1\,,
\end{equation}
exactly as the Kleinberg {\em upper limit} for this case.

The dominant $-1$ term on the lhs of (\ref{D1c}) might also be cancelled if $D(l)=1-g(l)$, where $g(l)$ vanishes as $l\to\infty$.  Substituting this ansatz in (\ref{D1c}) we find $g(l)\sim l^{2-\a}$, and integration of $D(l)$ yields
\begin{equation}
\label{a>2}
T\sim L+{\cal O}(L^{3-\a})\,,\qquad \a>2\,.
\end{equation}
In this case, the condition $\a>2$ prevents the correction term from growing faster than linearly.

\begin{figure}[t]
\includegraphics[width=0.45\textwidth]{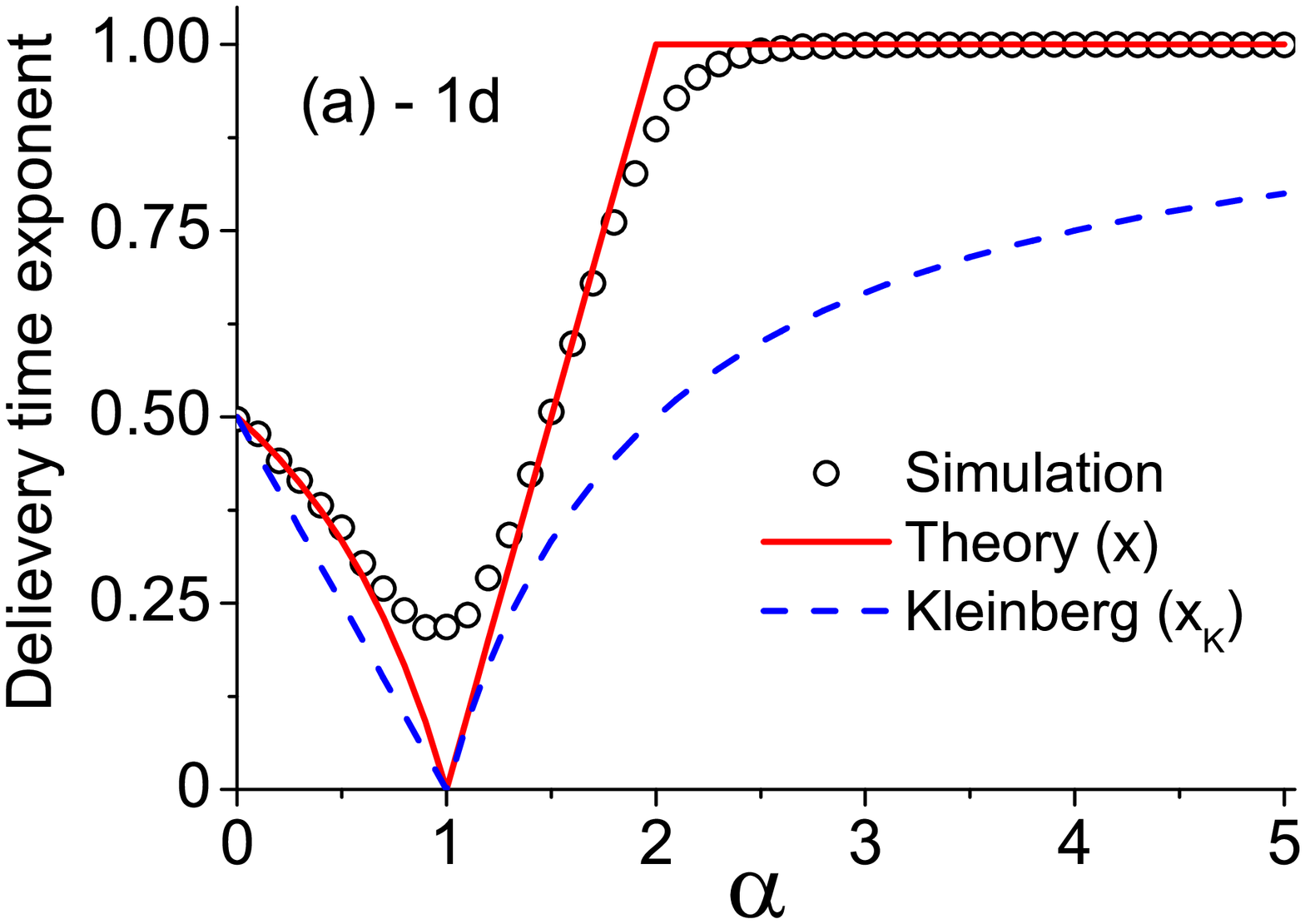}
\vskip-0.3in
\includegraphics[width=0.45\textwidth]{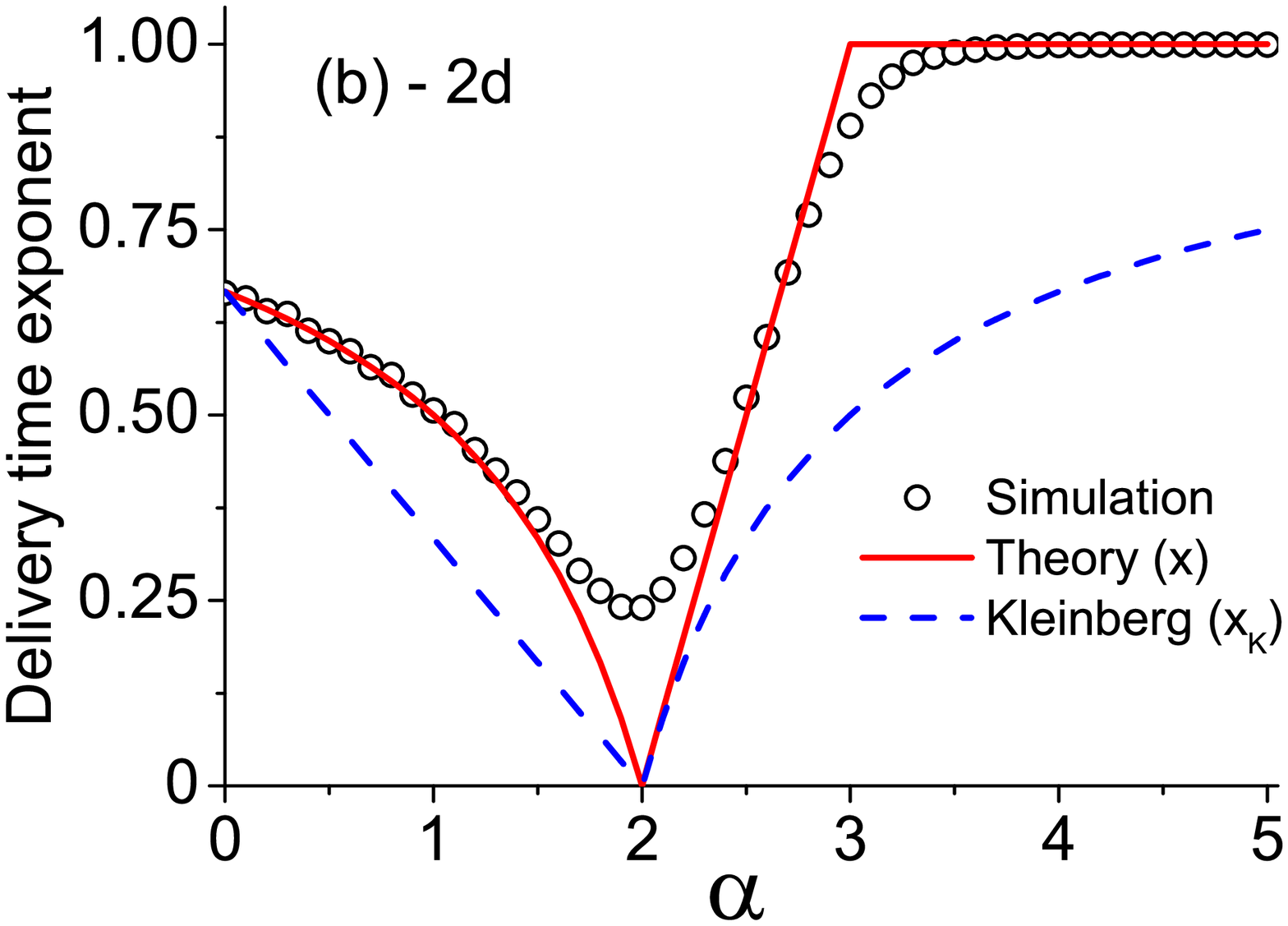}
\caption{(Color online) Delivery time exponent, $x$, as a function of $\a$ in (a)~$d=1$ and (b)~$d=2$ dimensions. Results from Eq.~(\ref{x}) (solid lines) are compared to the Kleinberg bounds (broken lines)
and to simulations (symbols). }
\label{x-alpha}
\end{figure}

In Fig.~\ref{x-alpha}a we compare our various results for $x(\a)$ in one dimension to computer simulations~\cite{remark2} and to the 
Kleinberg bounds, $x_K(\a)$.
For $\a\to2$ we expect logarithmic terms, as the main contribution and the correction term in~(\ref{a>2}) approach then the same power-law.  The logarithmic behavior makes it very difficult to extract reliable estimates from simulations for the 
exponent $x(\a)$ near $\a=1$ and~2.

The foregoing results, argued for one-dimensional lattices, are easily generalized to higher dimensions:
the space dimensionality, $d$, enters in the various integrals (or sums) through the Jacobian for $d$-dimensional integration.
Making the necessary adjustments, we find
\begin{equation}
\label{x}
T\sim L^x\,,\qquad x=
\begin{cases}
\frac{d-\a}{d+1-\a} & 0\leq\a<d\,,\\
\a-d & d<\a<d+1\,,\\
1 & \a>d+1\,.
\end{cases}
\end{equation}
Once again,  these results agree with the Kleinberg bounds $x\geq x_K$.  Eq.~(\ref{x}) fits simulations 
results for $d=2,3,4$ nicely (we did not test higher dimensions).  Results for $d=2$ are shown in Fig.~\ref{x-alpha}b.

Returning now to the probability distribution, $P(n;l)$, a standard way to tackle Eq.~(\ref{Pnl}) is through
the generating function
\[
\P(z;l)\equiv\sum_{n=0}^{\infty}P(n;l)z^n\,.
\]
The ``fugacity" $z$ may be interpreted as the probability to complete a single step successfully.  In that case
$P(z;l)$ is the total probability to complete the delivery successfully (to a target at distance $l$).
We defer a more detailed study of $P(n;l)$ for future work and focus for the moment on the conditional average of the delivery time, $T_{z,l}$, that is, the average delivery time conditioned upon successful arrival at the target:
\begin{equation}
\label{Tzl}
T_{z,l}=\frac{\sum_{n=0}nP(n;l)z^n}{\sum_{n=0}P(n;l)z^n}=z\frac{\partial \P(z;l)/\partial z}{\P(z;l)}\,.
\end{equation}

\begin{figure}[t]
\includegraphics[width=0.48\textwidth]{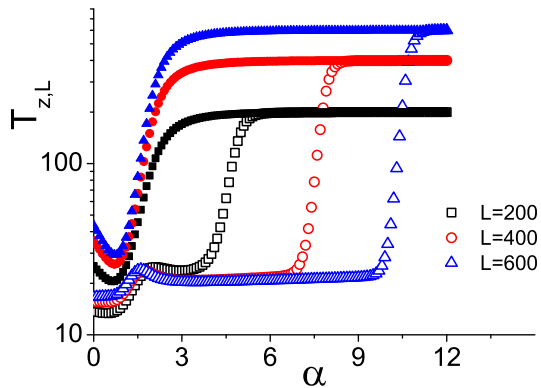}
\caption{(Color online) Conditional delivery time, $T_{z,l}$,  as a function of $\a$, for $L=200$ (squares),
400 (circles), and 600 (triangles).  Results for $z=0.9$ (empty symbols) are compared to perfect transmission
(solid symbols).}
\label{Tzl-alpha}
\end{figure}

Multypling~(\ref{Pnl}) by $z^n$ and summing over $n$, minding the boundary conditions, we obtain
\begin{equation}
\label{Pz}
\begin{split}
\P(z;l)=A&\sum_{k=1}^{2l-1}k^{-\a}z\P(z;|l-k|)\\
&+\Big(1-A\sum_{k=1}^{2l-1}k^{-\a}\Big)z\P(z;l-1)\,,
\end{split}
\end{equation}
 Taking the derivative of this equation with respect to $z$, and writing 
$\partial \P(z;l)/\partial z\equiv \F(z;l)$, we have
\begin{equation}
\label{Fz} 
\begin{split}
\F&(z;l)=A\sum_{k=1}^{2l-1}k^{-\a}[z\F(z;|l-k|)+\P(z;|l-k|)]\\
&+\Big(1-A\sum_{k=1}^{2l-1}k^{-\a}\Big)[z\F(z;l-1)+\P(z;l-1)]\,.
\end{split}
\end{equation}

We are now ready to address the conditional average, at least numerically.  Fixing the value of $z$,  we solve Eqs.~(\ref{Pz}) and (\ref{Fz}) recursively, up to $l=L$,  and use Eq.~(\ref{Tzl}) to compute the average.
Typical results are shown in Fig.~\ref{Tzl-alpha}.  For a fixed distance $L$ the delivery time is remarkably small throughout a wide range of long-contact exponents, $\a\lesssim \a_*(z;L)$, and saturates rapidly, $T_{z,L}\sim L$, as soon as $\a$ exceeds $\a_*$.  Thus, it seems that in the presence of losses small-world behavior does {\em not} imply a particular value of the long-contact exponent.  If anything,  there appears to be a small local {\em maximum} (barely perceptible in the figure) around a value of $\alpha$ that creeps towards $d$ as $L$ increases.  These results seem relevant to the Milgram experiment, where it was estimated
that $z\approx0.75$~\cite{white}.

In summary, we have found the actual delivery time exponent in the Kleinberg navigation model, for the 
non small-world cases of $\alpha\neq d$, and we have confirmed that $T\sim\ln^2L$ is the actual scaling (not just a bound) for the special case of $\a=d$. 
We have also introduced master equations that allow one to study the Kleinberg model analytically in greater generality,
and have looked briefly into the delivery time in the case of imperfect transmission, when there is a finite
probability for the message to get lost along its way.  The distribution of delivery times and the analytical treatment of imperfect transmission are important problems left open to further inquiry.

%
%

\acknowledgements

We thank L.~M.~Glasser and O.~Gromenko for many useful discussions. We are grateful to the Israel Science Foundation, to the Adams Fellowship Program of the Israel Academy of Sciences and Humanities (SC), and to the NSF, award PHY-0555312 (DbA), for partial support of this work.


\begin{thebibliography}{99}
\bibitem{remark2}
We actually used  $p_{i,i+k}\propto \int_k^{k+1}m^{-\a}\,dm$, instead of~(\ref{pijA}), for the ease of simulations.  This does not affect the asymptotic behavior of the search model.
\end{thebibliography}
\end{document}